\documentclass[a4paper,fleqn,usenatbib]{mnras}

\usepackage{newtxtext,newtxmath}
\usepackage[T1]{fontenc}
\usepackage{ae,aecompl}

\usepackage{graphicx}
\usepackage{amsmath}
\usepackage{amssymb}

\newcommand{\Msun}{\>{\rm M_{\odot}}}
\newcommand\degrees{^\circ}

\def\aj{AJ}
\def\apj{ApJ}
\def\apjl{ApJ}
\def\apjs{ApJS}
\def\aap{A\&A}
\def\mnras{MNRAS}
\def\pasp{PASP} 

\title[Galaxy tilting in the era of {\it Gaia}]{The tilting rate of the Milky Way's disc}

\author[S. W. F. Earp et al.]{Samuel W. F. Earp$^{1}$\thanks{E-mail: swfearp@gmail.com},
Victor P. Debattista$^{1}$,
Andrea V. Macci\`{o}$^{2,3}$, 
\newauthor David R. Cole$^{4}$\\
  $^1$ Jeremiah Horrocks Institute, University of Central Lancashire, Preston, PR1 2HE, UK \\
$^2$ New York University Abu Dhabi, PO Box 129188, Saadiyat Island, Abu Dhabi, UAE \\
$^3$ Max-Planck-Insitute for Astronomy, K\"{o}nigstuhl 17, D-69117 Heidelberg, Germany \\
$^4$ Rudolf Peierls Centre for Theoretical Physics, University of Oxford, Keble Road, Oxford, OX1 3NP, UK \\
}

\date{Accepted 2017 May 9. Received 2017 May 9; in original form 2016 October 13}

\pubyear{2017}

\begin{document}
\label{firstpage}
\pagerange{\pageref{firstpage}--\pageref{lastpage}}
\maketitle

\begin{abstract}
We present tilting rates for galaxies comparable to the Milky Way (MW) in a $\Lambda$ cold dark matter cosmological 
hydrodynamical simulation, and compare these with the predicted tilting rate detection limit of the {\it Gaia} 
satellite $0.28\degrees$Gyr$^{-1}$. We first identify galaxies with mass comparable to the MW 
($9 \times 10^{11} \le M_{200} \le 1.2 \times 10^{12} \Msun $) and 
consider the tilting rates between $z=0.3$ and $0$. This sample yields a tilting rate of $7.6\degrees \pm 4.5\degrees$Gyr$^{-1}$. 
We constrain our sample further to exclude any galaxies that have high stellar accretion during the same time. 
We still find significant tilting, with an average rate of $6.3\degrees$Gyr$^{-1}$. Both subsamples tilt with rates 
significantly above {\it Gaia}'s predicted detection limit. We show that our sample of galaxies covers a wide 
range of environments, including some similar to the MW's.  We find galaxies in denser regions tilt with 
higher rates then galaxies in less dense regions. We also find correlations between the angular misalignment of 
the hot gas corona, and the tilting rate. {\it Gaia} is likely to be able to directly measure tilting in the MW. 
Such a detection will provide an important constraint on the environment of the MW, including the rate of 
gas cooling onto the disc, the shape and orientation of its dark matter halo, and the mass of the Large Magellanic 
Cloud. Conversely, failure to detect tilting may suggest the MW is in a very quiet configuration.
\end{abstract}




\section{Introduction}
\label{sec:intro}
Disc galaxies such as the Milky Way (MW) are rapidly rotating; the
orientation of their spin axis represents the integral of the angular
momentum accreted via gas, interactions with satellites or other
galaxies, and torques exerted on the disc by the dark matter halo
within which they reside.  Therefore directly observing disc
tilting at the present time provides clues to the nature of each of
these processes. The {\it Gaia} space astrometry mission may soon allow direct
measurement of the MW's disc tilting rate. Precision
measurements will enable the construction of stellar position
catalogues with accuracies of order $20 \mu$ as with respect to distant
quasars, which act as the measurement reference frame
\citep{perryman2001, lindegren2008}.  \cite{perryman2014} estimate that an
accuracy better than $1 \mu$as yr$^{-1}$ should be achieved in all the
inertial spin components of the {\it Gaia} reference frame, corresponding to $0.28 \degrees$Gyr$^{-1}$.

Galaxies tilt for a variety of reasons. The role of interactions in disc tilting has been studied extensively.
While major mergers destroy discs, smaller scale interactions  are less violent, and tilt disc galaxies.
\citet{huangcarlberg1997} showed that infalling satellites tilt discs
so that there is a preference for infalling satellites to merge in the
plane of the disc.  \citet{read2008} reached a similar conclusion. \cite{bettfrenk2012} investigated the effects of minor mergers and flybys on the orientation of spins of dark matter haloes of mass $(12.0 \le \log_{10} (M/\Msun) h^{-1} \le 12.5)$ at $z=0$. They found that the majority of these events only caused small changes in the angular momentum of the entire halo, with only 10.5 per cent of MW mass haloes experiencing changes in their angular momentum by more than $45 \degrees$ over the course of their lifetimes. However, the inner halo is not so stationary, with 47 per cent of inner haloes experiencing a large change in their angular momentum orientation of at least $45 \degrees$ during their lifetimes. \cite{bettfrenk2015} extended this study to include a broader range of halo masses $(10.5 \le \log_{10}  (M/\Msun) h^{-1} \le 15.5)$. They found that 35 per cent of haloes had experienced changes in orientation of at least $45 \degrees$, at some point in their lifetimes, without a major merger taking place.

In the MW, the most important ongoing
interaction is with the Large and Small Magellanic Clouds (LMC and
SMC).  The mass of the LMC is currently the subject of debate, with
mass estimates as high as $M_\mathrm{LMC} \sim 2 \times 10^{11}\Msun$
\citep{kallivayalil2013, gomez2015, penarrubia2015}, corresponding to
$\sim 20$ per cent of the mass of the MW.  Other estimates are significantly lower ($\sim 5 \times 10^9 \Msun$) \citep{alvesnelson2000,vandermarel2002}.
Thus the importance of the LMC on the orientation of the MW's disc spin cannot
yet be estimated well.

Another cause of disc tilting is torques from dark matter haloes.
In the $\Lambda$-cold dark matter ($\Lambda$CDM) paradigm, haloes grow hierarchically, becoming
triaxial \citep{bardeen1986, barnes1987, frenk1988,
  duninskicarlberg1991, jingsuto2002, bailinsteinmetz2005,
  allgood2006}.  These triaxial
haloes are themselves tilting \citep{moore2004}.  \cite{dubinski1992}
examined the effect of tidal shear on dark matter haloes; he found that in all 14 of his $(1 \text{---} 2)
\times 10^{12} \Msun$ haloes the major axis rotated uniformly around
the minor axis with a rotation rate in the range of $6 \degrees \text{---} 96
\degrees $Gyr$^{-1}$. Likewise \cite{bailinsteinmetz2004} measured
figure rotation in 288 of their 317 dark matter haloes, finding a tilting rate of $6.2\degrees$Gyr$^{-1}$ with  
a log-normal distribution having $\sigma = 0.58 \degrees$Gyr$^{-1}$. \cite{bryancress2007} found that $63$ per cent of the 115 haloes they
considered exhibited significant figure rotation, with an average
pattern speed of $13.8 \degrees \; h \;$Gyr$^{-1}$.

The figure rotation of triaxial haloes leads to time varying torques on
discs.  \cite{debattista2015} showed that a stellar disc, lacking gas,
within a triaxial halo aligns its spin axis with the minor axis of the
halo. Even when
perturbed by a satellite the disc settles back to this alignment.
Thus a tilting halo will drag a disc along with it.
\cite{yurinspringel2015} inserted live stellar discs into eight, MW-sized, high-resolution dark matter haloes from
the AQUARIUS simulation. They found typical tilting rates of $5\degrees
\text{---} 6 \degrees$Gyr$^{-1}$, comparable with halo tilting rates.  While no
direct evidence of tilting haloes exists, tidal torques exerted on a
stellar disc by a rotating dark matter halo have been explored as a
possible cause for warps \citep{dubinskikuijken1995, dubinskichakrabarty2009}
and as a driving mechanism for spiral structure in dark matter-dominated
galaxies \citep{bureau1999}.

Galaxies such as the MW are generally thought to be surrounded
by hot gas coronae, with masses greater than the stellar disc
itself \citep[e.g.][]{spitzer1956, white1978, savageboer1979, white1991, dahlem1997,  wang2001,fukugita2006}. The quiescent cooling of this
hot gas then sustains star formation over a long time
\citep{fall1980, brook2004, keres2005, robertson2006, brooks2009}.  However,
the angular momentum of coronae is usually misaligned with that of their
embedded stellar disc \citep{vandenbosch2002, roskar2010}. This contributes misaligned angular momentum to
the disc, causing its orientation to change.
\citet{debattista2015} showed that under these circumstances, the
orientation of the disc spin is determined by a balance between the
torques from the triaxial dark matter halo, and the net inflow of angular
momentum via cooling gas.  As a result, star forming galaxies are
generally misaligned with the main planes of their dark matter haloes
\citep{sales2004, brainerd2005, agustsson2006, yang2006,
  azzaro2007,faltenbacher2007,wang2008a,wang2010, nierenberg2011, li2013}. \citet{debattista2013} argued for just
such an orientation in the MW, by noting that the best fitting
models for the Sagittarius Stream \citep{law2009, lawmajewski2010,
  degwidrow2013} require the disc spin to be along the halo's
intermediate axis, an orientation they showed is extremely unstable. 
\citet{debattista2013} therefore, argued that the modelling assumption
of the disc residing in one of the symmetry planes must be violated.
While this is indirect evidence, stacking of external galaxies has
shown that the distribution of satellites around blue galaxies tends
to be isotropic, contrary to what is seen around red galaxies \citep{sales2004,brainerd2005,yang2006,wang2008a,nierenberg2011,wang2014,dong2014}.

In summary in the MW, the disc may be tilting for a variety of reasons.  As a
first step towards understanding the tilting of the MW, in this paper, we
measure the tilting rates of MW-like galaxies in a $\Lambda$CDM cosmological
simulation. We compare the tilting rates of these discs to the observational
limit of {\it Gaia} to establish whether tilting of this nature would be
detectable. In Section \ref{sec:simulation}, we describe the cosmological
simulation. Then in Section \ref{sec:samples}, we describe the samples of
galaxies selected on the basis of virial mass, merger history and total satellite mass. In Section
\ref{sec:analysis}, we describe the methods we use to calculate the tilting
rates. Section \ref{sec:results}, presents the results, and provides a
comparison with the observational limit of {\it Gaia} for a variety of different 
local configurations and environments.  We present our conclusions in Section \ref{sec:discussion}, showing that even 
galaxies in quiet systems tilt at a rate that would be detectable by {\it Gaia}.


\section{Numerical Simulation}
\label{sec:simulation}

{The simulation we use here was performed with {\sc gasoline}, a
multi-stepping, parallel, tree code with smoothed particle hydrodynamics (SPH) 
 \citep{wadsley2004}. The version of {\sc gasoline} used for this work includes radiative and Compton
cooling for a primordial mixture of hydrogen and helium. The star formation
algorithm is based on a Jeans instability criterion \citep{katz1992}, but
simplified so that gas particles satisfying constant density, and temperature
thresholds in convergent flows spawn star particles at a rate proportional to
the local dynamical time \citep[see ][]{stinson2006}. The star formation
efficiency was set to $0.05$ based on simulations of the MW that
satisfied the Schmidt--Kennicutt Law \citep{schmidt1959,kennicutt1998}, and we adopt a star formation
threshold of 0.1 particles per cubic centimetre. The code also includes
supernova feedback using the blast-wave formalism as described in
\cite{stinson2006}, and a UV background following \cite{haardtmadau1996}; see
\cite{governato2007} for a more detailed description.

We used as a starting simulation one of the cosmological cubes described in
\cite{maccio2008}, namely our box has a size of  $180$Mpc and contained
300$^3$ dark matter particles. This box was created using WMAP5 \citep{komatsu2009} initial conditions with
($h$, $\Omega_M$, $\Omega_L$, $\Omega_b$, $\sigma_8$ ) $=$
($0.72$, $0.258$, $0.742$, $0.0438$, $0.796$) and was run with the code {\sc pkdgrav}
as detailed in \cite{maccio2008}.

From this simulation we selected at $z=0$ a volume of about (25 Mpc)$^3$
with  the requirement of not containing any haloes with a virial mass
above $5\times10^{12} \Msun$. For this purpose we use the halo catalogue
from  \cite{maccio2008} which was generated using a Spherical Overdensity
halo finder algorithm. The choice of this particular mass threshold
is motivated by our interest in studying the properties
of galaxies with a total mass equal or lower than the MW.

We then traced back to the initial conditions the Lagrangian region defined by this
redshift-zero volume, making sure to obtain a continuous 
region (i.e. no holes) at the initial redshift ($z=99$).
Finally, we used the standard  zoom-in technique to enhance the resolution
of the dark matter particles in the selected region  by a factor of $10^3$, and adding 
baryons (gas particles) with the same high resolution.
As a final result, this high resolution region contains more than $10^8$ 
particles, and reaches a mass resolution of $6.6 \times 10^6 $  and
$1.1 \times 10^6 \Msun$ for dark matter and gas, respectively, with a
gravitational softening length of $1.24$ kpc for dark matter and $0.5$ kpc for gas.

We then used the {\sc gasoline} code described above to evolve these
new high resolution initial conditions from $z=99$ to $0$ taking
into account gas cooling, star formation and feedback in a self consistent way.
To generate the catalogue of virially bound haloes we
use the grid based code {\sc amiga} Halo Finder
\citep{knollmannknebe2009} on the simulation outputs.


\section{The Samples}
\label{sec:samples}
We identify 182 haloes spanning the mass range
$9 \times 10^{10}$ \text{---} $4.4 \times10^{12}\Msun$. 
Of the 41 saved time steps during the time interval $z=0.3$ to $0$ we use a 
subset of ten time steps with an average 
separation of $\sim 0.37$Gyr to determine the stellar mass fractional 
growth rate, and to track the merger history of each galaxy. We calculate the tilting rate once for each 
galaxy, by measuring the angular momentum within 5 per cent of the virial radius at $z=0.3$ and $0$. 
 
\begin{figure}
  \begin{center}
    \includegraphics[width=\columnwidth]{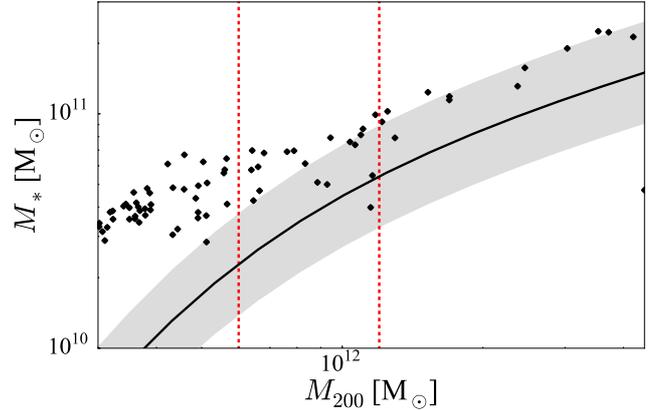}
    \caption{Stellar mass versus halo mass at redshift $z = 0$ for the most massive galaxies in the initial
    sample (black points). We measure the stellar mass within 5 per cent 
    of the virial radius ($r_{200}$),  where the mean interior density is 200
    times the critical density. For comparison the black line shows
    the $M_{*} \text{--} M_{200}$ relation of \citet{kravtsov2014} derived using halo
    abundance matching. The grey shaded region shows the scatter in this
    relation. The red dashed lines illustrate the bounds that subsample A
    lies within.}
    \label{fig:abundance}
  \end{center}
\end{figure}

Subsample A contains haloes within a specified mass range comparable to the MW.
The motivation for this mass cut is two-fold. First, we are interested in galaxies
with similar halo mass as the MW. Secondly, we wish to choose galaxies where
the mass of the dark matter halo, and the stellar mass are in good agreement with 
abundance matching results. We impose an upper limit of $M_{200} \leq 1.2 \times 10^{12} \Msun$
in order to constrain the sample to a mass range that is comparable with the virial
mass of the MW \citep{klypin2002}. We also find that above this limit the full sample is
dominated by ellipticals that, due to their evolutionary history, generally
have lower specific angular momentum. Fig. \ref{fig:abundance} compares
the halo mass--stellar mass relation of the full sample with the relation derived by the
abundance matching method of \citet{kravtsov2014}.  This figure shows that galaxies residing within 
haloes of mass $M_{200}\ge 9 \times 10^{11}\Msun$ in the
simulation match this relationship well. Lower mass haloes, however, have an excess
stellar mass. Therefore, we use this mass as a lower limit for subsample A. Implementing the mass range $1.2 \times 10^{12}   \ge M_{200} \ge 9 \times 10^{11}\Msun$ leaves $19$ galaxies in
subsample A.

Subsample B has the same mass constraint with two
added limits: one on the change in stellar mass to remove galaxies that
have undergone mergers above a certain mass ratio, and the second on the ratio of galaxy mass to total satellite mass.
First, we observe the evolution of the galaxies by visual inspection to construct a catalogue of galaxies that
have not undergone mergers between $z=0.3$ and $0$. By comparing this catalogue to the full sample we
can constrain the rate of change in stellar mass such that above this limit 
the sample is dominated by galaxies that have undergone mergers. Fig. \ref{fig:mergercut}
shows the distribution of fractional growth rates for galaxies that do not undergo mergers and the full sample.
 From Fig. \ref{fig:mergercut} we set an upper limit on the maximum stellar mass fractional growth 
rate of $0.16 $Gyr$^{-1}$ within 5 per cent of the virial radius,
under which galaxies have not undergone significant minor or major mergers.
We construct subsample B from subsample A with the added constraint that $\Delta M_* / (\langle M_* \rangle \Delta t)$ must fall below this value. 
Two galaxies that fell below this limit were observed undergoing a minor merger, however, the maximum stellar mass accreted was roughly one per cent that of the central galaxy so they are included in subsample B.
Lastly, we stipulate that the total satellite stellar mass must be less than 40 per cent that of the central galaxy at every time step. 
To measure the total satellite mass we subtract the total stellar mass within $0.1 r_{200}$, where $r_{200}$ is the virial radius, from the total stellar mass inside $r_{200}$. This leaves us  with just seven galaxies in subsample B. 

\begin{figure}
  \begin{center}
    \includegraphics[width=0.99\columnwidth]{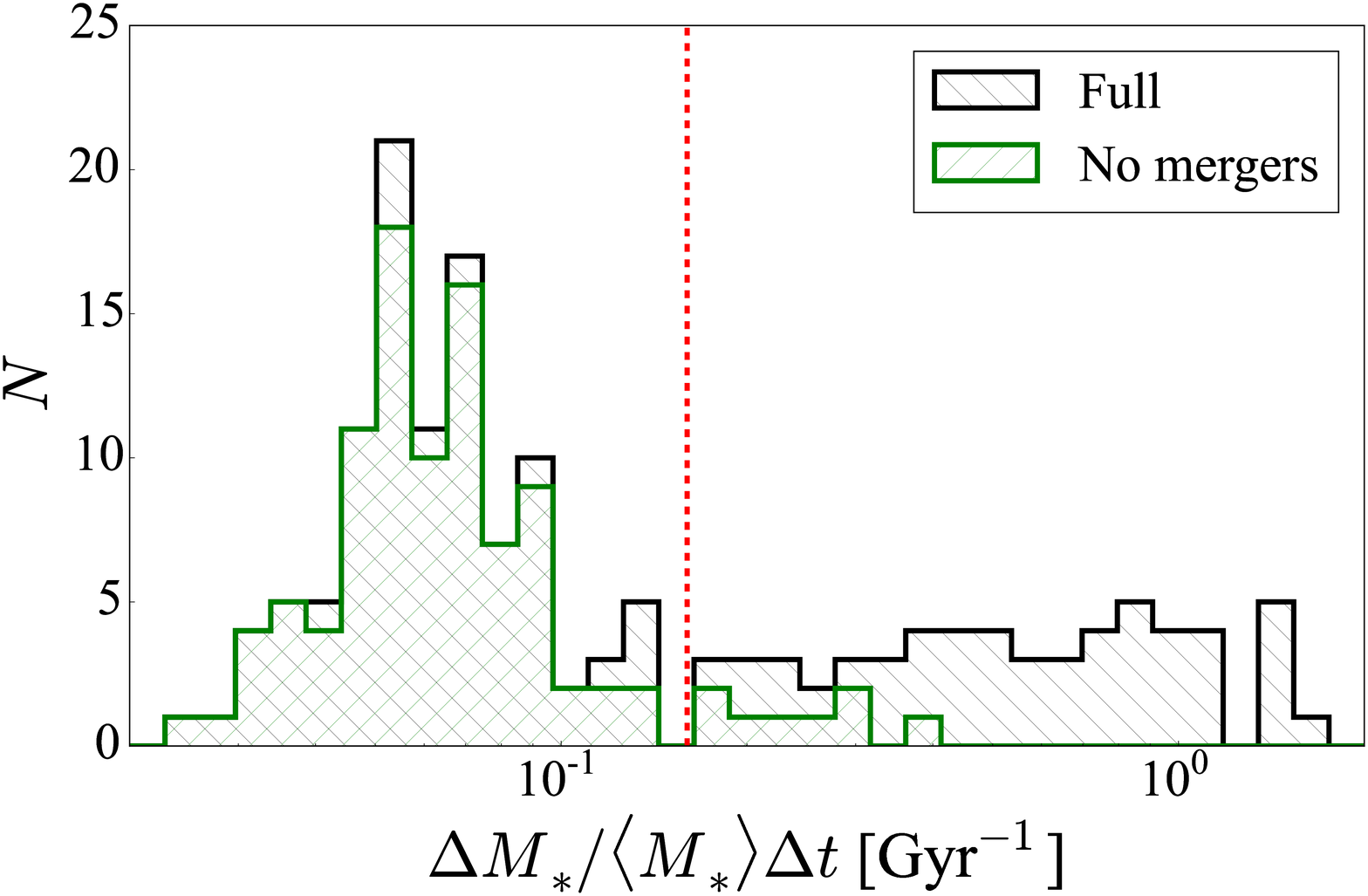}
    \caption{Distribution of the maximum rate of change in stellar mass normalized by average stellar mass between $t_i$ and $t_{i+1}$ within 5 per cent of the virial radius. The black histogram shows the rates for the full sample of 182 haloes. The green histogram shows the distribution of rates for galaxies that did not undergo any mergers between $z=0.3$, and $0$. The red dashed line indicates the upper limit we impose on subsample B.}
    \label{fig:mergercut}
  \end{center}
\end{figure}


\section{Analysis}
\label{sec:analysis}
To derive accurate tilting rates, we 
first find the kinematic centres of the galaxies.  We adopt the position of
the lowest potential dark matter particle as our kinematic centre. 
We verify that this method is reliable by computing the kinematic
centre using an iterative shrinking sphere method.  Starting with a sphere of $200$ kpc, we 
iterate centring on the centre of mass and halving the radius each step to a final value of $\sim10$ pc.  Using the lowest 
potential dark matter particle, we are able to obtain kinematic centres for our entire sample of 182 galaxies.

We then measure the angular momentum of the galaxy by summing the angular
momentum of each star particle within $R < 0.05 r_{200}$. This radius is
selected to include the disc of the galaxy, but exclude any warps. \cite{briggs1990} 
found that warps become detectable within the Holmberg radius ($R_{\rm{H}o}$). For a 
typical virial radius of $\sim 200$ kpc, we would expect a Holmberg radius of $\sim 15$ kpc, $5$ kpc greater than the radius we would 
consider for $0.05r_{200} \sim 10$ kpc.
We also select this radius to avoid selecting just 
the bulges of our galaxies, which tend to have lower specific angular momentum.

In order to determine the uncertainty in the tilting rates, we measure the difference
in the direction of the angular momentum vector at different radii. We measure the angular momentum
at seven linearly spaced radii spanning $0.01 < R/r_{200} < 0.04$. We then use the average angular discrepancy between
the vectors as the error $(\sigma)$ on the measurement of the angular momentum vector, and hence on the tilting rate. 
For each of these errors we assign a weight $w$ such that $w = 1/\sigma^2$, which will be used 
in the calculation of the mean and standard deviation of each subsample.


\section{Results}
\label{sec:results}

\subsection{Tilting rates}
First we consider subsample A, i.e. galaxies with virial mass comparable to the MW's, within the range 
$9\times10^{11} \le M_{200}\le1.2\times10^{12}\Msun$. We measure the tilting rate once, 
between the two time steps $z=0.3$ and $0$. This subsample 
tilts with a mean rate of $7.6\degrees$Gyr$^{-1}$,
and a standard deviation of $4.5\degrees$Gyr$^{-1}$, well above the average error 
for this subsample of just $0.05\degrees$Gyr$^{-1}$. All 19 of the galaxies in this 
subsample exhibit significant tilting above  {\it Gaia}'s detection limit of $0.28 \degrees$Gyr$^{-1}$ \citep{perryman2014}.

Next we consider subsample B, i.e. the galaxies with similar mass to the MW, that have low fractional stellar mass change from $z=0.3$ to $z=0$, and have a maximum 
total satellite mass of $40$ per cent that of the central galaxy.  Fig. \ref{fig:tiltwitherr} shows tilting rates for sample B
versus the ratio of stellar mass to satellite stellar mass. Each data point in this figure corresponds to a tilting
rate of a single galaxy, with the mass ratio measured at $z=0$. The green squares show only the five galaxies that 
were not observed to undergo any mergers since $z=0.3$, while the black squares were the two galaxies that did undergo a minor 
merger within the same time.
The tilting rates of this subsample have an average of $6.3\degrees$Gyr$^{-1}$, with a standard deviation of 
$6.5\degrees$Gyr$^{-1}$, well above the average uncertainty of $0.13\degrees$Gyr$^{-1}$. 
This subsample also tilts with a rate well above {\it Gaia}'s detection limit.

\begin{figure}
  \begin{center}
    \includegraphics[width=0.99\columnwidth]{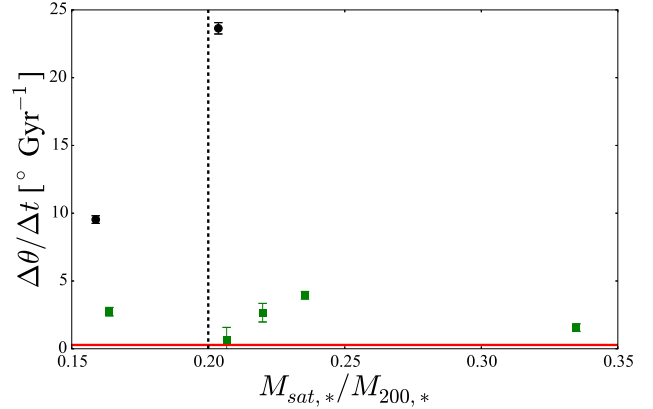}
    \caption{Tilting rate versus the present day fraction of satellite to galaxy stellar mass ($M_{\rm{sat},*}/M_{200,*} $) for subsample B, i.e. galaxies that
    have mass comparable to the MW, and have low fraction stellar mass change between $z=0.3$ and $0$. The (green) squares represent the galaxies that were observed to not have undergone any mergers since $z=0.3$, the (black) circles show the galaxies that undergo a minor merger. The black dashed line shows recent estimates of the mass ratio of the LMC relative to the MW \citep{kallivayalil2013, gomez2015, penarrubia2015}. The red horizontal line is {\it Gaia}'s predicted detection limit \citep{perryman2014}.}
    \label{fig:tiltwitherr}
  \end{center}
\end{figure}

\subsection{Environmental dependence}
To determine if there is any dependence between the tilting rates of galaxies and their local environment, 
we compare the tilting rates of the galaxies with their normalized local density. We calculate the density 
within various radii centred on each galaxy, and then normalize by the critical density at $z=0$. 
Fig. \ref{fig:tiltvsnormdens} shows the distribution of densities for spheres with radii $3,4,5$ and $6$ Mpc. We find 
that for large radii ($5$ and $6$ Mpc) that there is a strong correlation for subsample A with $p$ values of 
$0.8$ for both, although, for smaller radii ($3$ and $4$ Mpc), the correlation weakens, with $p$ values of $0.2$ and 
$0.6$ respectively. When we consider subsample B the correlations are enhanced, with $p$ values of $0.7, 0.95, 0.97$ and $0.96$ for radii $3,4,5$ and $6$Mpc, respectively.

\begin{figure*}
  \begin{center}
    \includegraphics[width=\textwidth]{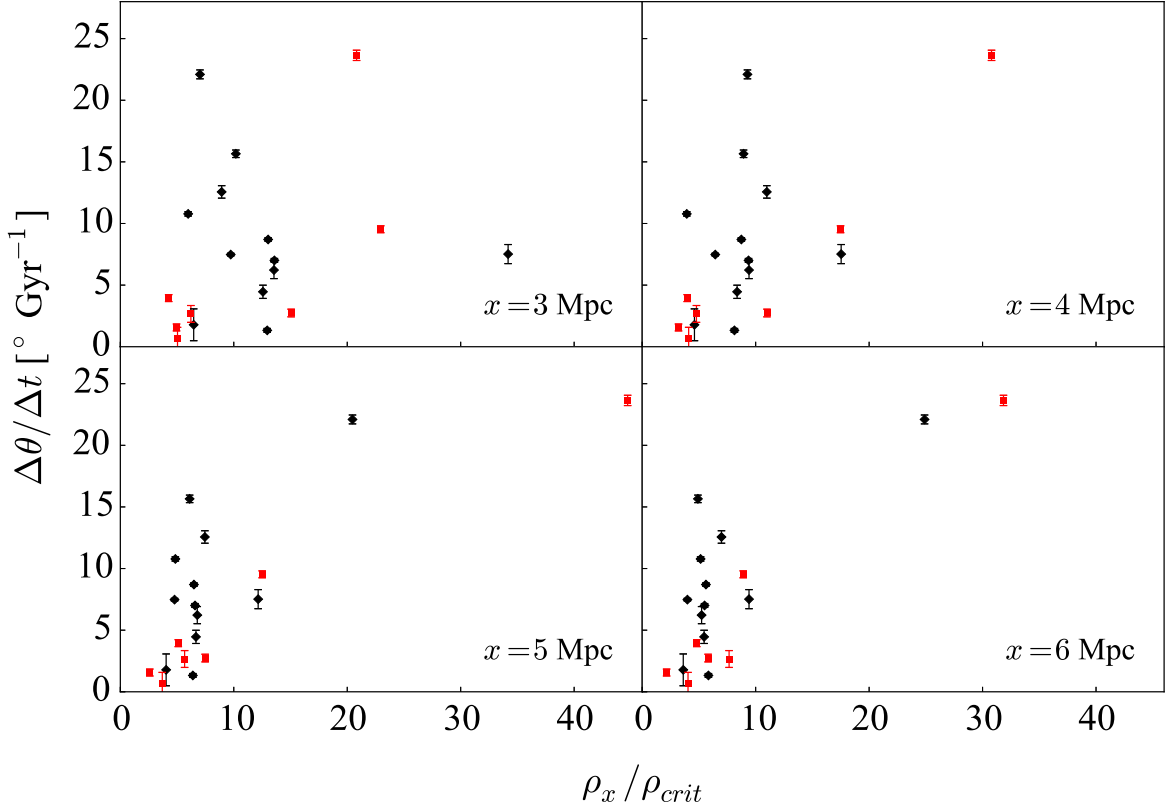}
    \caption{Tilting rate versus the local density within a sphere of radius $x$ at redshift $z=0$. In all panels, the (black)  diamonds represent galaxies in subsample A with masses comparable to the MW,and the (red) squares show galaxies in subsample B with comparable mass and undergoing no interactions since $z=0.3$. We measure correlation coefficients for each panel $x = 3,4,5$ and $6$ Mpc of $p=0.2, 0.6, 0.8$ and $0.8$, respectively, for all points, while for subsample B, we find $p$ values of $0.7, 0.95, 0.98$ and $0.97$, respectively.}
    \label{fig:tiltvsnormdens}
  \end{center}
\end{figure*}

The MW has a close massive neighbour M31 within $1$ Mpc. We compare the tilting rates with the distance $D$ 
to the nearest massive ($M_* > 9 \times 10^{11}\Msun$) galaxy in subsample A. Fig. \ref{fig:tiltvsdistance} shows the tilting rate versus $D$; 
galaxies in subsample A span a range of $D$, including some with very close neighbours and 
some very isolated. We see no relation between $D$ and the tilting rate. Considering 
galaxies in subsample B we do find a weak anti-correlation. One of our galaxies does appear to be tilting 
extremely fast without a close neighbour.

\begin{figure}
  \begin{center}
    \includegraphics[width=0.99\columnwidth]{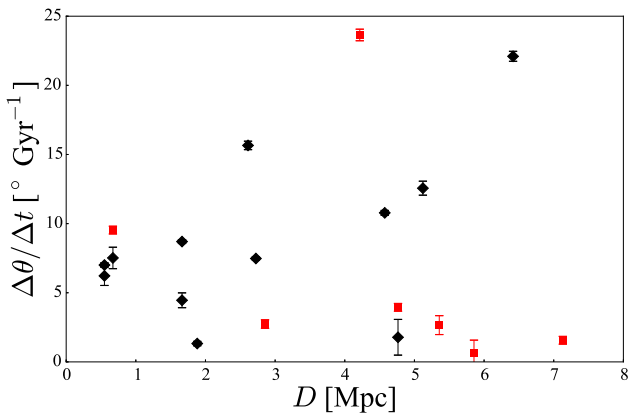}
    \caption{Tilting rate versus distance $D$ to nearest galaxy with comparable mass to the MW measured at $z=0$. The (black) diamonds represent subsample A and the (red) squares show subsample B. We find a correlation coefficient of $p= -0.05$ for subsample A and $p = -0.3$ for subsample B.}
    \label{fig:tiltvsdistance}
  \end{center}
\end{figure}

\subsection{Dependence on gas}
The angular momenta of the hot gas corona surrounding a galaxy and of the disc are not generally aligned. As the gas corona continually 
feeds cool gas to the disc, this misalignment causes gas being accreted to change the angular momentum of the disc. 
To investigate this effect on the tilting rate, we define the hot gas corona in two different ways. In the first, we choose all gas with a temperature 
$T>5 \times 10^4$K, and in the second, we choose all gas between two spherical shells of radii $0.2r_{200}$ and $r_{200}$. 
The angular momentum calculated from each definition is in good agreement, with $p = 0.99$. We compare the tilting rates of the hot gas corona to the tilting rate of the disc for both of these methods. 
Fig. \ref{fig:dthecordthe} shows that for both methods of defining the corona, there is no correlation 
between the angular momentum tilting rate of the corona and disc for MW mass galaxies. Even when we consider 
subsample B, we find no correlation for both methods.

\begin{figure*}
  \begin{center}
    \includegraphics[width=0.99\columnwidth]{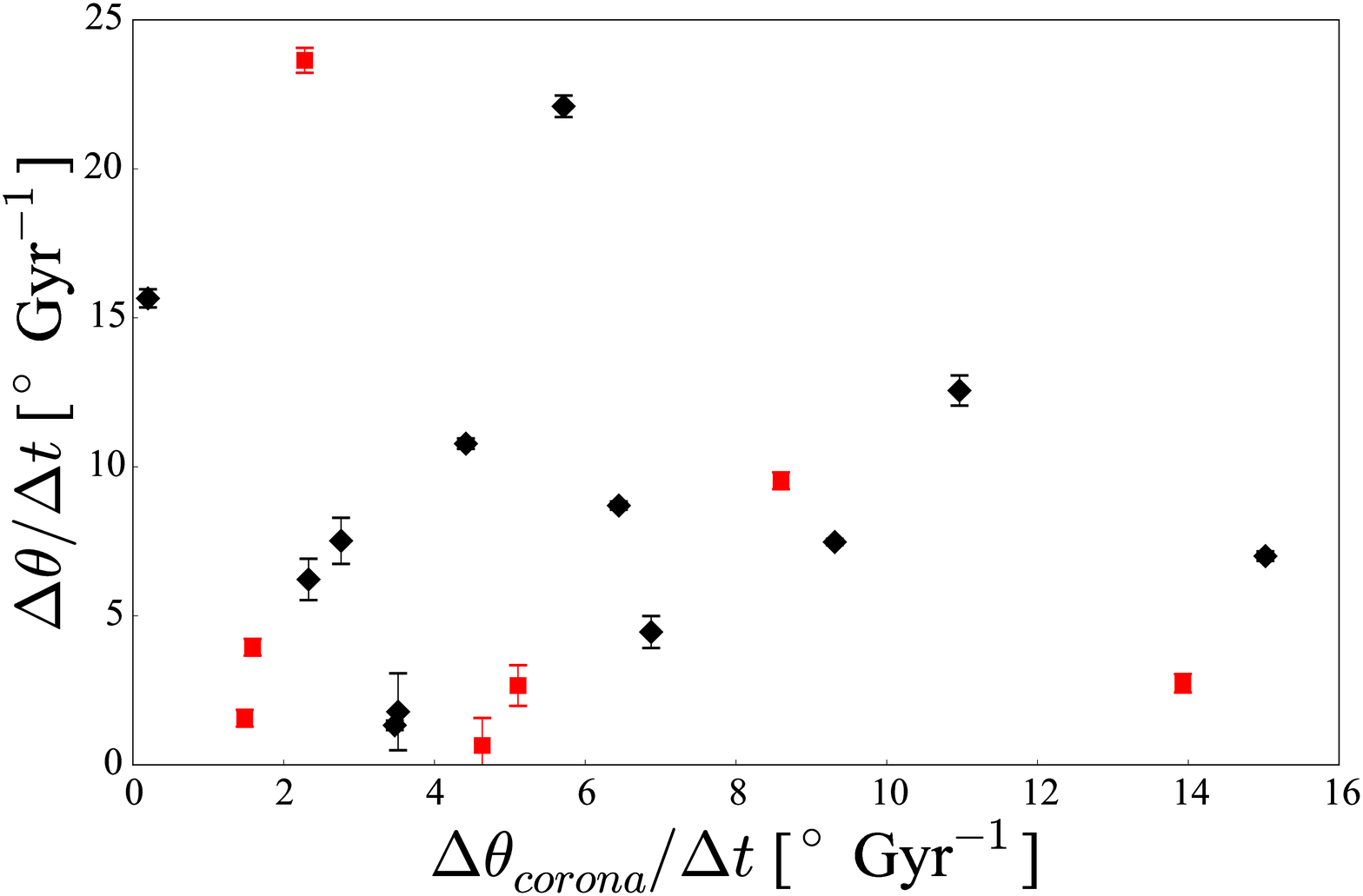} \includegraphics[width=0.99\columnwidth]{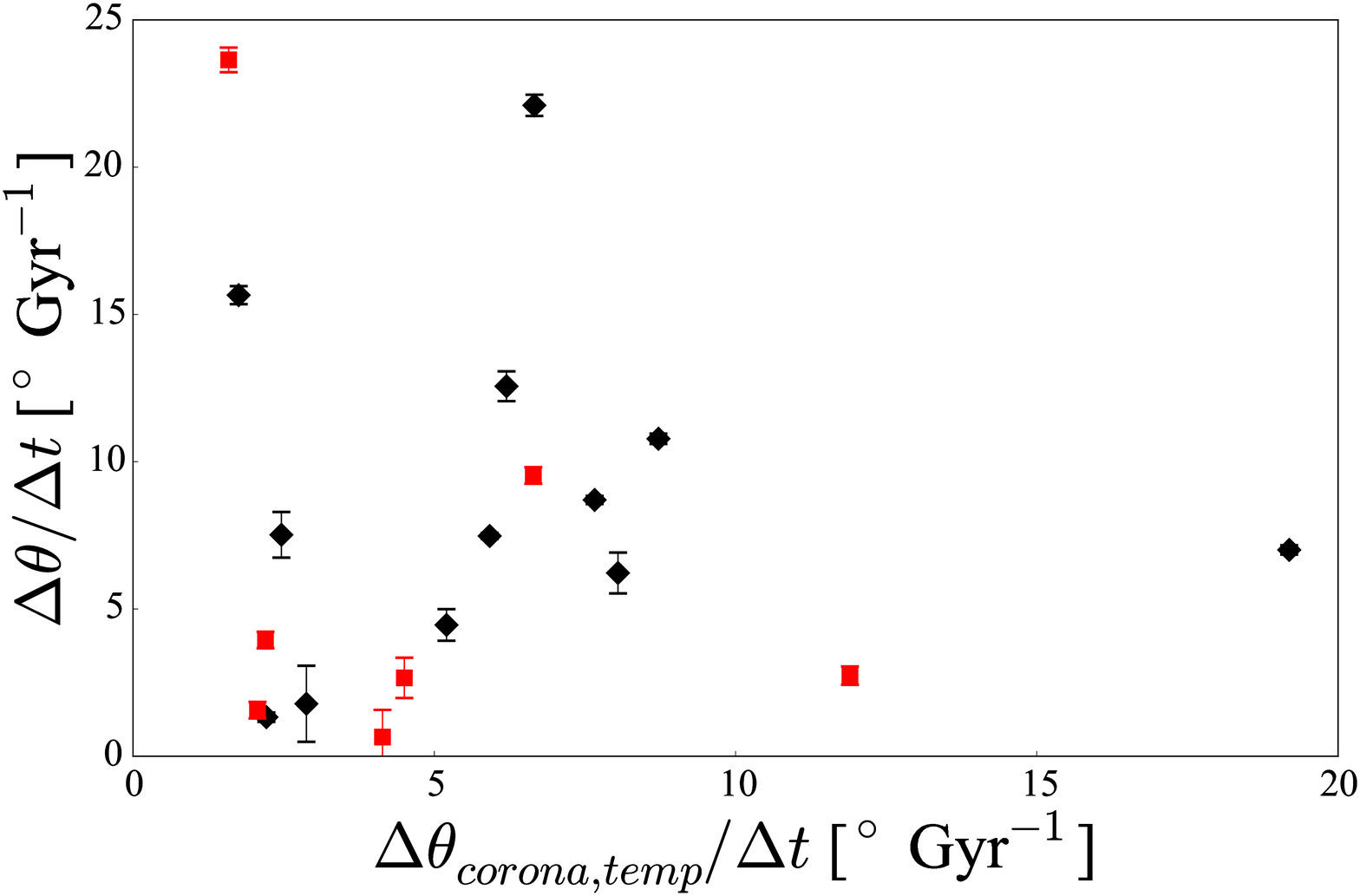}
    \caption{Left-hand panel: The tilting rate of the stellar disc versus the tilting rate of the corona, defined as all gas between radii $0.2r_{200}$ and $r_{200}$. We find no  correlation for subsample A (black diamonds) with a coefficient of $p = -0.08$. For subsample B (red squares) we find no correlation with $p = -0.18$. Right-hand panel: The tilting rate of the stellar disc versus the tilting rate of the corona, defined as all gas with a temperature $T > 5 \times 10^4$ K. We find no significant correlation for subsample A (black diamonds), with $p = -0.035$. Similarly for subsample B (red squares) we find no correlation, with $p = -0.27$.}
    \label{fig:dthecordthe}
  \end{center}
\end{figure*}

Next we compare the tilting rates of the disc to the angular misalignment between the disc and the hot 
gas corona for both methods of defining the hot gas corona. Figure \ref{fig:corthedthe} shows the relation 
between the tilting rates of discs and the angular 
misalignment of the hot gas corona and disc angular momentum. We find a weak correlation with $p$ values of $0.4$ and $0.5$ 
for both methods, respectively, for subsample A. 
However, for subsample B, the correlation strengthens considerably for 
both methods with $p$ values of $0.86$ and $0.87$. 

The large-scale structure (LSS) may influence the flow of gas into the halo and subsequently the misalignment between 
the stellar and coronal angular momentum. When we compare the misalignment of the hot gas corona from the 
stellar disc with the normalized local density, we find similar correlations as those we found in Fig. 
\ref{fig:tiltvsnormdens}. Therefore it is not possible to determine from this simulation if the effect of the 
environment directly governs the tilting of the galaxy, or if the LSS affects the tilting via its effect 
on the coronal angular momentum, as seems likely. \citet{debattista2015} found that galaxies 
lacking gas generally aligned with the minor axis of their halo. However, when gas is allowed to cool 
on to the disc, the orientation can be more arbitrary. For both of our subsamples, we find that 
galaxies with higher star formation generally tilt with higher rates. These results favour the 
gas driven tilting scenario. 

When we compare the angular momentum misalignment between the disc and the gas corona with the local density, we find similar correlations as those in Fig. \ref{fig:tiltvsnormdens}.  Thus the mechanism by which the LSS affects the disc's tilting rate is unclear. The LSS may torque the disc directly, or it may influence the flow of gas into the halo, driving the misalignment between the stellar and coronal angular momentum, which in turn drives the tilting \citep[e.g. ][]{debattista2015}.  One possible clue comes from comparing the tilting and the star formation rate. For both of our subsamples, we find that galaxies with a higher star formation rate generally tilt faster, suggesting that it is the delivery of misaligned angular momentum through gas that dominates the tilting.

\begin{figure*}
  \begin{center}
    \includegraphics[width=0.99\columnwidth]{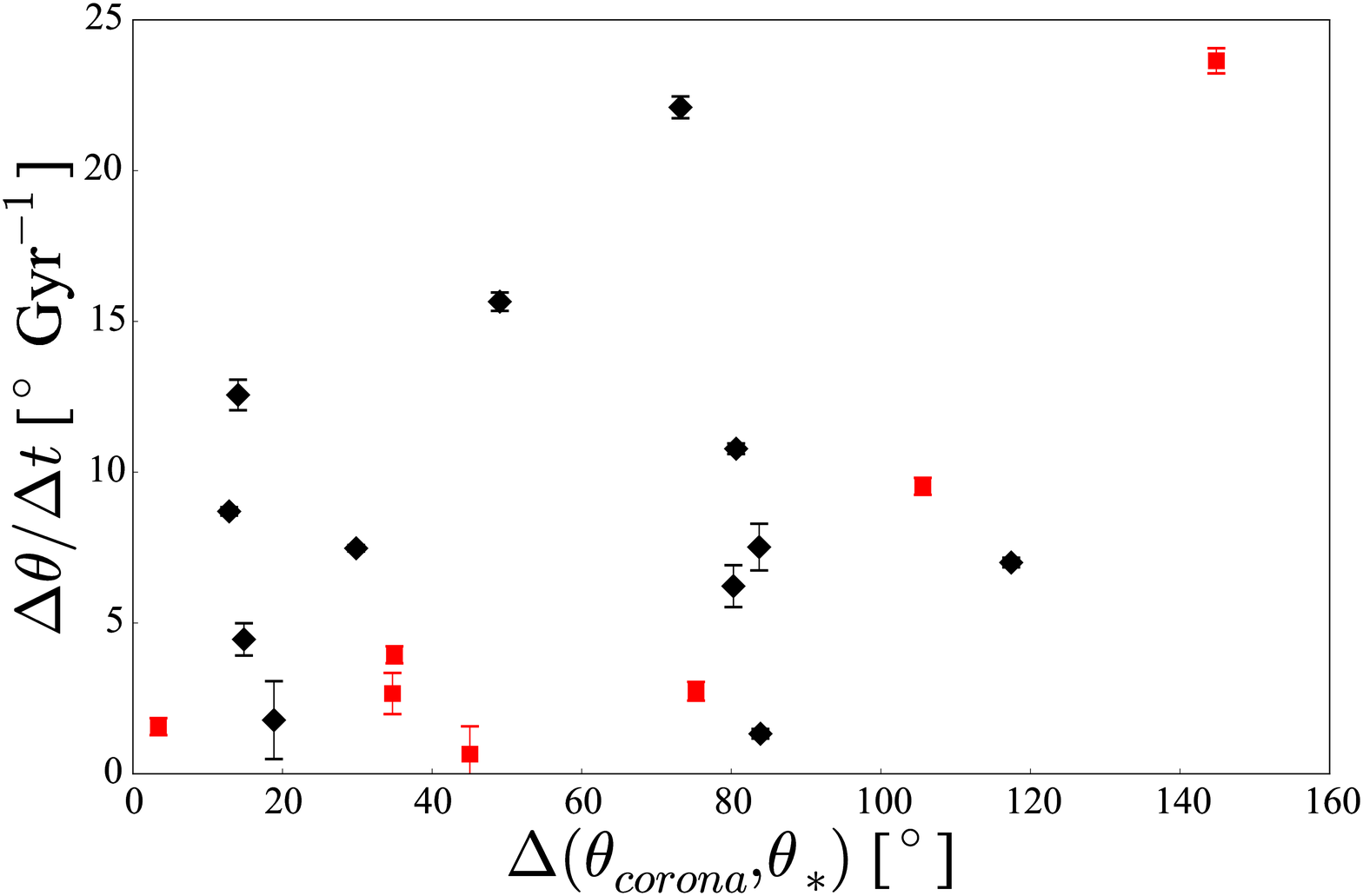} \includegraphics[width=0.99\columnwidth]{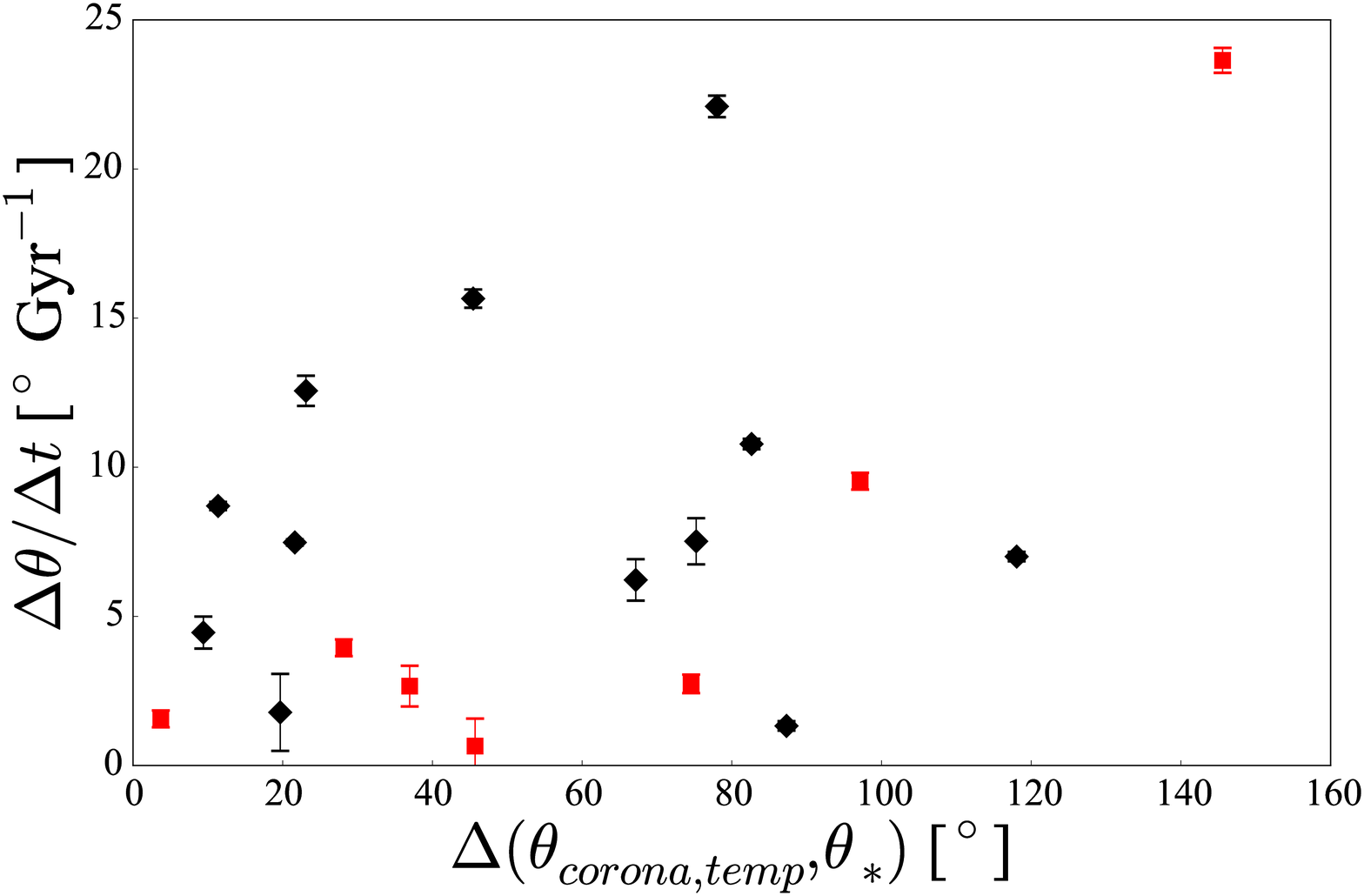}
    \caption{Left-hand panel: The tilting rate of the stellar disc versus the angular difference in angular momentum orientation between the stellar disc and the hot gas corona, defined as all gas between radii $0.2r_{200}$ and $r_{200}$. We find a weak correlation for subsample A (black diamonds), with $p = 0.4$, and a strong correlation for subsample B (red squares), with $p = 0.86$. Right-hand panel: The tilting rate of the stellar disc versus the difference in orientation between the stellar disc and the hot gas corona, defined as all gas with temperature $T > 5 \times 10^4$ K. We find a weak correlation for subsample A (black  diamonds) with $p = 0.5$, and a strong correlation for subsample B (red squares) with $p = 0.87$.}
    \label{fig:corthedthe}
  \end{center}
\end{figure*}


\section{Discussion and Conclusions}
\label{sec:discussion}
When we consider galaxies with halo masses comparable to the MW 
(subsample A), we find significant tilting with an error weighted mean rate of $ 7.6 \degrees$Gyr$^{-1}$
 and a standard deviation of $4.5 \degrees$Gyr$^{-1}$. The entire subsample displays 
significant tilting with rates higher than the detection limit of {\it Gaia}. We further restrict to a sample with low relative stellar accretion, 
and a maximum stellar mass fraction in satellites of $40$ per cent (subsample B), finding a lower mean 
tilting rate of $6.2\degrees$Gyr$^{-1}$, with a range from $0.65$ to $24.6\degrees$Gyr$^{-1}$.

A variety of processes may drive the change in angular momentum 
that we have measured. Interactions with other galaxies are the most violent processes 
changing the angular momentum of discs drastically over a short period.
However, we have found that even when we exclude strong interactions we still measure significant tilting
above the detection limit of {\it Gaia}. Therefore, we must turn to secular processes such as 
halo torques and the accretion of misaligned cold gas on to the disc to explain 
the entire phenomena of disc tilting. 

We investigated the effect of the local environment on 
the tilting rate of the disc. Comparing the local density against the tilting rate, we find that 
the tilting rate does not correlate with the normalized local density within $3$Mpc for subsample A.
However, for subsample B, we do find a correlation.
When we consider larger radii, we find a correlation between the tilting rate 
and the local environment for both subsamples. Galaxies in denser regions generally tilt at higher rates  
than galaxies in lower density regions, irrespective of the galaxies stellar mass accretion. 

The MW has a very close, similar mass, neighbour M31. In order to compare 
to the MW's configuration, we measured the distance to the nearest massive 
galaxy and determined the correlation with the tilting rates. We find almost no correlation 
for subsample A; however, for subsample B, we do find very weak anti-correlation. This suggests that the local configuration is 
unlikely to be a large contributing factor when the disc is accreting significant mass. Our 
sample contains galaxies in similar configurations to the MW with companion galaxies within 
a few hundred kpc; these galaxies exhibit tilting rates similar to more isolated galaxies.

To determine the effect of misaligned gas accreting from the hot gas corona, 
we measured both the tilting rate of the hot gas 
corona and the angular misalignment between the stars and the corona. We find no correlation between the tilting rates of the two 
different components for either subsample. We also compared 
the tilting rate of the disc to the angular momentum misalignment between the two components: 
For subsample A, there is a weak correlation, which becomes stronger for subsample B. We also find a correlation between the misalignment of the disc and coronal angular momentum and the LSS. Thus, the LSS may directly affect the tilting rate via torques, or indirectly by influencing the flow of gas into the halo. For both subsamples, galaxies with higher star formation tilt faster, perhaps indicating that the role of the LSS is in driving the misaligned gas.
We conclude that the angular momentum misalignment between the corona and disc is an important, possibly dominant, driver of disc tilting.

In this paper, we have measured the tilting rates for a wide variety of galaxies of similar mass 
to the MW, in various configurations, some similar to the local configuration of the MW. 
Every configuration yielded a tilting rate above the {\it Gaia} limit and should be detectable.
Confirmation of a tilting disc would have important consequences for
understanding the evolution of the MW.  For example, the tilt of the
disc will make the potential seen by the Sagittarius Stream time varying.
Conversely failure to detect tilting may suggest the MW is in an unexpectedly quiet configuration.

\section*{Acknowledgements}
SWFE would like to thank Dominic Bowman for useful conversations.
VPD is supported by the
Science and Technology Funding Council Consolidated grant \#ST/M000877/1. VPD acknowledges the support of the Pauli Center
for Theoretical Studies, which is supported by the Swiss National
Science Foundation (SNF), the University of Z\"{u}rich, and ETH
Z\"{u}rich. Simulations were performed at the Rechen Zentrum of the Max Planck Society in Garching (RZG) on the {\sc theo} and {\sc hydra} machines.



\bibliographystyle{mnras}
\bibliography{earp+2016}


\bsp
\label{lastpage}
\end{document}